# Design of a variable-Mach-number waverider by the osculating-curved-cone method using a rational distribution function and incorporating the equilibrium-gas model

Mengyu Wang[a], Yi Duan[b], Qin Li[a,1], Luying Lin[a], Chuan Tian[b]

a. School of Aerospace Engineering, Xiamen University, China, 361102
b. Science and Technology on Space Physics Laboratory, China Academy of Launch Vehicle Technology, Beijing 100076, China

**Abstract:** When a waverider flies at hypersonic speed, the thermodynamic properties of the surrounding gas change because of the rapid increase in temperature, so it is reasonable to consider real-gas effects in the vehicle design. In addition, a hypersonic waverider usually travels at varying speed during flight, and deviating from the default speed designed in terms of a constant Mach number often creates difficulties in preserving the expected performance. Therefore, research on the design of variable-Mach-number waveriders considering equilibrium-gas effects is important for applications. In this paper, a design method for a variable-Mach-number osculating-curved-cone waverider (VMOCCW) considering equilibrium-gas effects is introduced, then the influences of different gas models on the waverider design are studied by taking a VMOCCW designed with a linear Mach-number distribution as an example. Furthermore, a new Mach-number distribution method is proposed by using a parameterized rational function, which is combined with different gas models to achieve VMOCCW design. For comparison, waveriders designed with quadratic concave and convex increasing functions are also selected for comparison of their layouts and aerodynamic performances under design and off-design conditions. The results show that waveriders designed with the equilibrium-gas model exhibit differences in geometric features (e.g., volume and volumetric efficiency) and aerodynamic characteristics (e.g., lift-to-drag ratio and pitching moment coefficient) compared to those designed with the ideal-gas model. Specifically, waveriders designed with a rational function for the *Ma* distribution have a wing-like structure, and overall they have more-balanced geometric and aerodynamic characteristics than those designed with quadratic concave and convex functions.

# 1 Introduction

As a highlight of aerospace technology, hypersonic aircraft are extremely fast over very long distances and a wide range of altitude. To achieve these advantages, the aircraft usually must have a high lift-to-drag ratio but also a high volumetric efficiency, and achieving both makes the layout design very challenging. As a cutting-edge vehicle that is currently being studied, the waverider is a hot topic in hypersonic layout design because of its characteristics of high lift, low drag, and high lift-to-drag ratio [1, 2], and research communities have conducted extensive research on its design theory, flow simulation, optimization techniques, and engineering applications. At present, waverider design methods mainly employ basic flows such as two-dimensional wedge flow [2], supersonic cone flow [3], and three-dimensional supersonic flow of cone–wedge combination [4], as well as the osculating method of supersonic axisymmetric flow [5]. Representative of the latter type is the osculating cone method proposed by Sobieczky et al. [5]. Differences aside, the overall process for designing a conventional waverider is as follows [6]: 1) determine the overall design parameters; 2) design the basic profiles; 3) compute the basic flow and derive the leading-edge layout of the waverider in terms of the basic flow, shock profile, and geometric characteristics; 4) perform streamline tracing in both the free stream and basic flow to generate the upper and lower surfaces of the waverider given the leading-edge layout, followed by closing the bottom surface to fulfill the design.

---

[1] Corresponding author, email: qin-li@vip.tom.com

During the hypersonic flight of a waverider, the effects of a real gas [as opposed to an ideal gas (IG)] at high temperature are inevitable, with physical and chemical reactions such as dissociation, recombination, and ionization likely to occur among the gaseous components. As such, the thermodynamic properties of the gas differ significantly from those of an ideal one. When the Mach number ($Ma$) exceeds 10, real-gas effects become more obvious and have a significant impact on the longitudinal static stability and maneuverability of the waverider [7]. If these effects are ignored, then a waverider designed using the IG model may encounter poor longitudinal trim characteristics when in flight [8]. As reported in the literature, computational fluid dynamics (CFD) is the usual means of considering real-gas effects in waverider design, with the advantages of high solution accuracy and the ability to design arbitrary shapes and accommodate subsonic flow scenarios. However, CFD is less efficient than the method of characteristics (MOC), whose efficiency and precision make it suitable for the rapid design of hypersonic aircraft. Obviously, enhancing the fidelity of the physical models in the MOC (e.g., the gas model) helps to improve its accuracy, with practical importance for engineering applications. Because it is computationally expensive to simulate real-gas effects when using chemical nonequilibrium models, the equilibrium gas (EG) model can be used efficiently to describe the effects in engineering applications, especially when the time for the chemical reactions to reach equilibrium is shorter than the flow characteristic time. There are two main types of EG model: 1) those that involve solving the component equations along with the flow control equations (e.g., the equilibrium-constant method, the minimum-free-energy method, etc.); 2) the method of fitting curves of experimental data. For the latter, Tannehill and Mohling [9] proposed their version of curve fitting for air, and Tannehill and Mugge [10] improved it using piecewise functions. Srinivasan et al. improved its accuracy further [11] and proposed a fitting relation for the transport coefficient in the context of viscosity [12]. Overall, the curve-fitting method is easily programmed and efficiently implemented and so is used widely in computations of chemical equilibrium flows. To the best of the present authors' knowledge, there has been little research to date on EG effects in the MOC and its use in waverider design.

In addition to real-gas effects, waveriders inevitably undergo variations of speed and airspace during takeoff, cruise, and landing as well as maneuvering. At the design Mach number, the shock wave of the waverider is confined approximately to the compression surface with little overflow at its edges, thereby achieving a good lift-to-drag ratio. However, in an off-design state, there may be spillover at the edges, resulting in reduced aerodynamic performance. Therefore, efforts have been made to develop waverider design theory in order to improve the aerodynamic performance in off-design states and achieve better adaptation to a wide speed range. Currently, there are at least two methods for doing so; one is represented by the "vortex lift" waverider, which offers improved low-speed performance; the other involves designing waveriders with a wide range of high speeds, e.g., the combined-and-spliced waverider, the variable-Mach-number waverider (VMW), and the multistage-compression waverider. Of these, the VMW is the result of a new design method for wide-speed-range waveriders, the main idea being to design a waverider based on multiple Mach-number solutions corresponding to the basic flow. By increasing the discrete $Ma$ from two to three or more, in other words, devising its distribution along the spanwise direction of the waverider, $Ma$ can be distributed smoothly in different regions, via which is expected an aircraft with satisfactory aerodynamic performance in a wide speed range. In this regard, Zhang et al. [13] first proposed the VMW concept, then using the same idea, Li et al. [14] put the design into practice and analyzed the performances of the derived waveriders. Referencing the cone-derived waverider with variable $Ma$, Zhao et al. [15, 16] developed a design method for an osculating-cone VMW by combining the idea of osculating cones with variable $Ma$, and they analyzed the effects of several $Ma$ distributions on the layout and aerodynamic performance of the waverider. Liu et al. [17] further studied VMW design by using the theory of osculating flow. Meng et al. [18] performed design with a constant conical plane shape (the shock profile is straight and the shock angle in each osculating plane is the same) considering variable $Ma$, and they discussed its wide-speed-range characteristics in the hypersonic range. All of those studies gave improved waverider aerodynamic performance under off-design conditions to different degrees, hopefully accommodating a flight environment with a full speed range and large airspace [19].

Despite the above progress on VMWs, the following points warrant further investigation. 1) A waverider operating under hypersonic conditions experiences real-gas effects, but the relevant design methods have rarely been considered to date (except in [20]). Therefore, a lack of investigation and understanding of these effects exists on the design layout and aerodynamic performance. 2) Although the discussion in [17] addressed how characteristics of the $Ma$ distribution such as linearity/nonlinearity, monotonicity, and concavity/convexity influence the design of an osculating-cone VMW, the types that were not covered mean that further studies are necessary to derive a waverider with the balanced high lift-to-drag ratio and volumetric efficiency needed in engineering applications. In particular, further research into optimizing the $Ma$ distribution function is warranted. 3) Low-speed aircraft have a wing-like layout, and it is worth studying whether similar structures can be incorporated naturally into hypersonic VMWs.

Given the above issues and noting previous work on the MOC considering EG effects [20], proposed herein is a method for designing a variable-Mach-number osculating-curved-cone waverider (VMOCCW) using a newly devised rational function (RF) for the $Ma$ distribution and incorporating the EG model as well as the canonical IG one. Using the derived waveriders, the effects of different gas models and design methods are then compared and analyzed. Section 2 introduces the method for designing a VMOCCW with EG effects considered, including solving for the basic equilibrium flow and presenting numerical verifications. In Section 3, the VMOCCW derived using a linear $Ma$ distribution is taken as an example to discuss the geometric and aerodynamic differences of the waverider due to the different gas models. In Section 4, the $Ma$ distributions for the osculating-cone VMW are reviewed, then a new parametric RF is proposed for the $Ma$ distribution, which is combined with the different gas models to design a VMOCCW. A layout with a wing-like structure is obtained, and its configuration and aerodynamic performance is studied under design and off-design conditions. Finally, Section 5 concludes the paper.

## 2 Design of VMOCCW considering equilibrium-gas effects

### 2.1 Review of VMOCCW design methods

In 1990, Sobieczky [5] first proposed the theory of osculating-cone design, in which the shock profile at the bottom of the waverider is defined freely according to requirements instead of being limited to an arc (for a cone-derived waverider) or a straight line (for a wedge-derived waverider), thereby expanding considerably the design space and application range of the waverider [6]. Then based on Sobieczky's theory, He and Ni [21] proposed a method for designing an osculating-curved-cone waverider via the following concrete design process. Given the design parameters, the curved-cone generatrix equation, the flow capture curve (FCT), and the inlet capture curve (ICC), to be solved for are the streamlines on the lower surface and bottom profile. At this time, the ICC can be regarded as an envelope curve enclosing countless arc segments, the generation of which arises from a conical shock wave intersecting with the bottom section. As exemplified in Fig. 1(a), taking a point $P_1$ on the ICC, the circle that passes through $P_1$ and is tangent to the ICC is the corresponding osculating circle (defined by the intersection between the basic shock wave and the bottom section). The osculating-circle center is point $O_1$, and the radius is the curvature radius of the ICC at $P_1$. The plane passing through line $O_1P_1$ and perpendicular to the bottom section is the osculating plane $AA_1$, as shown in Fig. 1(b). In osculating plane $AA_1$, point $P_2$ corresponds to the intersection of the osculating plane and the FCT; the curved-cone generatrix is shown as OBC, where the half-cone angle $\delta_1$ is to be determined. The generatrix consists of a straight line OB and a smooth curve segment BC, and both are continuously connected at point B. Segment BC is determined uniquely by the curve equation (e.g., cubic curve) together with the coordinates and tangent angles $\delta_1$, $\delta_2$ of the two endpoints. After determining the configuration of the curved cone, the MOC is used to derive the basic flow, which is then scaled in the osculating plane so that the shock point and axis point at the exit of the basic flow match with $P_1$ and $O_1$ in the osculating plane [21]. The horizontal projection of point $P_2$ on the basic shock wave generated by the curved cone yields the leading-edge point P. By tracing the streamline along the leading-edge point P in the curved-cone flow, the lower-surface streamline and corresponding bottom trailing-edge point $P_3$ are obtained. Connecting all trailing-edge points yields the bottom profile of the lower surface of the waverider, and all similar streamline sets constitute its lower surface.

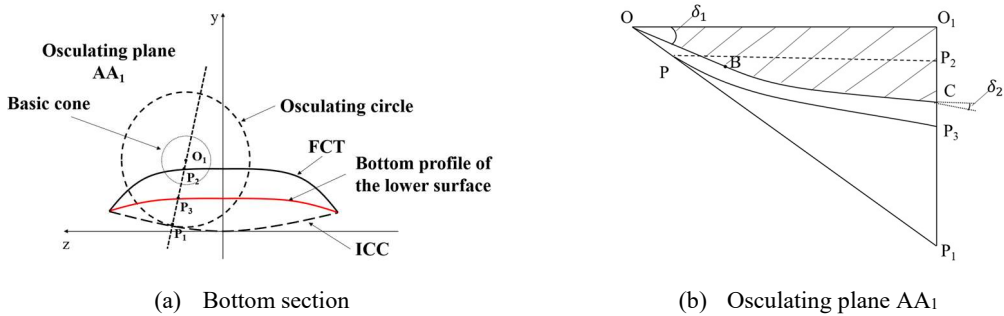

(a) Bottom section      (b) Osculating plane $AA_1$

**Fig. 1.** Schematics of an osculating axisymmetric waverider

The principle of VMOCCW design is basically the same as that of the conventional osculating-curved-cone method. The main difference is that varying the design *Ma* is specified in the incoming flow conditions of each osculating-plane flow field [17]. Referring to [14], the design steps are summarized as follows.

1) Given the range of the design $Ma$ or $[Ma_{min}, Ma_{max}]$, determine the discrete distribution of $Ma$ along the spanwise direction of the waverider. Taking a linear increasing distribution from the edge to the symmetric meridian plane of the waverider as an example, if $Ma$ is discretized uniformly into $n$ segments, then the corresponding distribution function is

$$Ma = Ma_{min} + \frac{(Ma_{max} - Ma_{min})}{n-1}(i-1) \quad (1 \leq i \leq n) \tag{1}$$

2) Given the design parameters including the design shock angle $\beta$, the curved-cone generatrix equation, the FCT, and the ICC, determine the cone configuration and the basic flow in each osculating plane. To ensure the smoothness of the leading edge of the waverider, it is recommended in [22] that each osculating plane should correspond to the same design shock angle. In detail, the ICC is first discretized into several points to derive the related osculating planes, then the design $Ma$ associated with each osculating plane is determined according to the $Ma$ distribution. Next, the cone half-angle is defined empirically via $\frac{1}{Ma(\beta - \delta_1)} = \frac{(\gamma+1) Ma \times \beta}{1 + \frac{\gamma-1}{2} Ma^2 \beta^2}$ [14], thereby defining the curved cone. Finally, the MOC is used to solve for the basic flow, which is then scaled in the corresponding osculation plane.

3) By horizontally projecting the intersection of the osculating plane and the FCT onto the corresponding basic shock surface, the leading-edge point is obtained that is affiliated with the osculating plane. In the curved-cone flow, by tracing the streamlines generated from the leading-edge points to the trailing-edge points of the bottom section, the streamline can be derived in each osculating plane, which is actually the intersection of the waverider layout with the osculating plane. The combination of the leading-edge points of each osculating plane forms the leading-edge curve, and that of the trailing-edge points forms the trailing-edge profile of the lower surface.

4) The streamlines in all the osculating planes form the lower surface of the waverider. The leading-edge profile and the bottom profile of the upper surface form the upper surface, while both bottom profiles of the upper surface and the lower surface form the bottom surface. The upper surface, the lower surface, and the bottom surface together comprise the VMOCCW

From the literature, the IG model is usually used in the above procedure to solve for the basic flow. Herein, we especially consider EG effects and apply the corresponding model in the VMOCCW design. Unlike with the traditional design method, it is expected that EG effects will affect the solutions for the shock waves and flows around the (curved) cone, which refers to the respective solution methods for the oblique shock relations, the Taylor–MacColl equations, and the MOC elements. The corresponding details are introduced briefly below.

## 2.2 Conical flow solution for equilibrium-gas flow

For EG flow, the relationship among the state variables is not the simple IG formula; instead, it must be determined via curve fitting or the equilibrium-constant method. Herein, the curve-fitting method by Tannehill and Mohling [9] is used to derive the EG solution. In particular, from the pressure $p$ and density $\rho$, the equivalent specific heat ratio $\tilde{\gamma}$ and temperature $T$ can be solved for via the fitting formulas $\tilde{\gamma} = \tilde{\gamma}(p, \rho)$ and $T = T(p, \rho)$, with each formula or fitted curve comprising piecewise functions with different value ranges. In concrete calculations, input $p$ and $\rho$ and compute $X = log_{10}(p/p_0)$ and $Y = log_{10}(\rho/\rho_0)$, then choose specific forms of the fitting formulas to derive the unknown thermodynamic variables and $\tilde{\gamma}$ based on $X$ and $Y$. Specifically, the curve-fitting formula for $\tilde{\gamma}$ is

$$\tilde{\gamma} = c_1 + c_2 Y + c_3 Z + c_4 YZ + c_5 Y^2 + c_6 Z^2 + c_7 Y^2 Z + c_8 YZ^2 + c_9 Y^3 + c_{10} Z^3 + \frac{(c_{11} + c_{12}Y + c_{13}Z + c_{14}YZ + c_{15}Y^2 + c_{16}Z^2 + c_{17}Y^2Z + c_{18}YZ^2 + c_{19}Y^3 + c_{20}Z^3)}{[1 \pm exp(c_{21} + c_{22}Y + c_{23}Z + c_{24}YZ)]} \quad (2)$$

where $Z = X - Y$, $\rho_0 = 1.292 kg/m^3$, $p_0 = 1.0134 \times 10^5 N/m^2$. The "$\pm$" choice in the denominator is determined by the value ranges of $X$ and $Y$; when "+" is chosen, the corresponding piecewise function is referred to as the odd function transition, otherwise it is the even function transition. See [9] for the details of $c_i$. For $T = T(p, \rho)$, the fitted curve is given by

$$log_{10}(T/T_0) = d_1 + d_2 Y + d_3 Z + d_4 YZ + d_5 Y^2 + d_6 Z^2 + d_7 Y^2 Z + d_8 YZ^2 + d_9 Y^3 + d_{10} Z^3 + \frac{d_{11} + d_{12}Y + d_{13}Z + d_{14}YZ + d_{15}Y^2 + d_{16}Z^2 + d_{17}Y^2Z + d_{18}YZ^2 + d_{19}Y^3 + d_{20}Z^3}{[1 \pm exp(d_{21} + d_{22}Y + d_{23}Z + d_{24}YZ)]} \quad (3)$$

where $T_0 = 273.15$ K (likewise, $T$ is in kelvin) and $Y$, $Z$, $p_0$, and $\rho_0$ are the same as those in Eq. (2). See [9] for the details of $d_i$. The sound speed $a$ and the specific enthalpy $h$ are derived from $\tilde{\gamma}$ and $T$ as

$$a = \sqrt{\tilde{\gamma} RT} \quad (4)$$

$$h = \left(\frac{p}{\rho}\right)\left(\frac{\tilde{\gamma}}{\tilde{\gamma}-1}\right) \quad (5)$$

To design the cone-derived waverider, it is necessary to establish the conical basic flow first, and a similar approach is used to generate the initial flow in basic-flow computations of the curved-cone waverider. The conical flow is derived by solving the Taylor–MacColl equations, and the corresponding flow model is sketched in Fig. 2, where $\theta_c$ is the half-cone angle, $r$ is the radius vector, $\theta$ is the angle between $r$ and the cone axis, and $V_r$ and $V_\theta$ are the velocity components along and perpendicular to $r$, respectively.

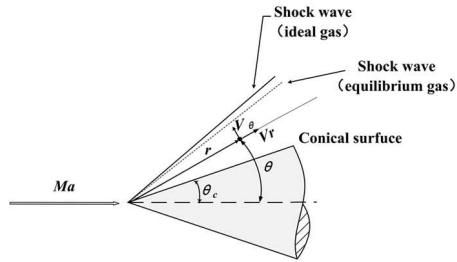

**Fig. 2**. Schematic of conical flow

For an EG, the Taylor–MacColl equations [23] are

$$\frac{dV_\theta}{d\theta} = \frac{a^2}{V_\theta^2 - a^2}\left(2V_r + V_\theta \cot\theta - \frac{V_r V_\theta^2}{a^2}\right) \quad (6)$$

$$\frac{dp}{d\theta} = -\frac{\rho V_\theta a^2}{V_\theta^2 - a^2}(V_r + V_\theta \cot\theta) \tag{7}$$

$$\frac{d\rho}{d\theta} = -\frac{\rho V_\theta}{V_\theta^2 - a^2}(V_r + V_\theta \cot\theta) \tag{8}$$

$$V_\theta = \frac{dV_r}{d\theta} \tag{9}$$

where Eq. (7) was proposed specifically in [20] to accommodate EG effects, and $a$ is computed by Eq. (4). In the case of an IG, Eqs. (6), (8), and (9) are applied, and $a$ is given by $a = \sqrt{\gamma RT}$ with $\gamma = 1.4$.

## 2.3 MOC for equilibrium-gas flow

In designing a curved-cone waverider with EG effects considered, the flow around the sharp cone is usually solved for via the Taylor–MacColl equations and then used as the initial-value line to compute the rest of the basic flow. Concretely, the solution at the 0.01 m wall location of the straight cone is chosen as the initial-value line for the MOC, via which the downstream solution can be achieved by advancing. As an accurate stepwise method for solving quasi-linear partial differential equations [24], the MOC is used to solve for inviscid supersonic/hypersonic internal and external flows. The following is a brief introduction.

In inviscid steady supersonic flow, the characteristic line passing through a certain point in the flow is indicated in Fig. 3. There are three characteristic lines through one point P in the flow: a streamline $C_0$, a left-running Mach line $C_+$, and a right-running Mach line $C_-$. The characteristic equations [24] along the streamline and the left-running and right-running Mach lines are

$$\left(\frac{dy}{dx}\right) = \lambda_0 = \frac{v}{u} \text{(along the streamline)} \tag{10}$$

$$\left(\frac{dy}{dx}\right)_\pm = \lambda_\pm = \tan(\theta \pm \alpha) \text{(along the } C_\pm) \tag{11}$$

and the corresponding compatibility equations are

$$\rho V dV + dp = 0 \text{(along the streamline)} \tag{12}$$

$$\frac{\sqrt{Ma^2 - 1}}{\rho V^2} dp_\pm + d\theta_\pm + \delta\left[\frac{\sin\theta dx_\pm}{yM\cos(\theta \pm \alpha)}\right] = 0 \text{(along the } C_\pm) \tag{13}$$

where the subscript "0" indicates a parameter along the streamline, and the subscript "$\pm$" indicates a parameter along the left-running or right-running Mach line; $\theta$ is the flow direction angle, $\alpha$ is the Mach angle, and $\delta = 0$ corresponds to planar flow while $\delta = 1$ corresponds to axisymmetric flow. The above characteristic and compatibility equations are consistent for an IG or EG. In the case of an IG, the compatibility equation along the streamline is

$$dp - a^2 d\rho = 0 \tag{14}$$

where $a$ is the local sound speed, and the relationship between pressure and density is that for an IG, i.e., $p = \rho RT$. In the case of an EG, the compatibility equation along the streamline is Eq. (4) [20].

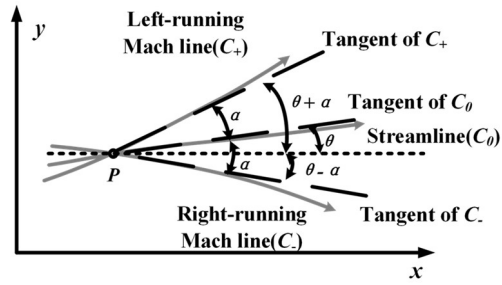

**Fig. 3.** Schematic of characteristic lines passing through point P

Another important part of the MOC is solving for the shock boundary element. For an IG, the oblique shock relation is used directly to obtain the post-shock variables. For an EG, the following governing equations are used:

$$\rho_1 u_1 = \rho_2 u_2 \tag{15}$$

$$p_1 + \rho_1 u_1^2 = p_2 + \rho_2 u_2^2 \tag{16}$$

$$h_1 + \frac{u_1^2}{2} = h_2 + \frac{u_2^2}{2} \tag{17}$$

where the subscript "1" indicates a wavefront variable and the subscript "2" indicates a post-shock variable. $\rho_2$ is given by $\rho_2 = \rho(p_2, h_2)$ from Eq. (5) combined with the Tannehill fitting formula, thereby closing Eqs. (15)–(17). Given $\rho_1$, $u_1$, $p_1$, and $h_1$ of the upstream flow, the ratios $\rho_2/\rho_1$, $u_2/u_1$, $p_2/p_1$, and $h_2/h_1$ are obtained. According to [25], the theoretical oblique shock angle of an EG is smaller than that of an IG.

### 2.4 Verification computations

To verify the correctness of the conical flow solution and the MOC considering EG effects, the conical flow and the curved conical flow were each computed. First, computations were performed for a conical flow field with zero angle of attack, an inflow Mach number of $Ma$ = 8–42, and a cone half-angle of $\theta_c = 46°$. The reference results are those given by Hudgins [26] with the same conditions. Fig. 4 shows the distributions of dimensionless wall pressure $p/p_\infty$, density $\rho/\rho_\infty$, and temperature $T/T_\infty$ with $Ma \times sin\theta_c$ at the heights of zero, 100,000 ft, and 200,000 ft, and the computed results (Comp) are compared with the reference results (Ref). The two sets of results agree well, which verifies the correctness of the methods in Section 2.2.

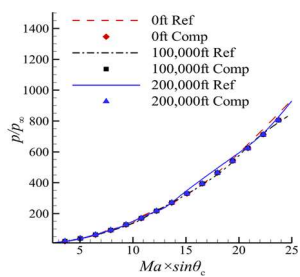
(a) Wall pressure

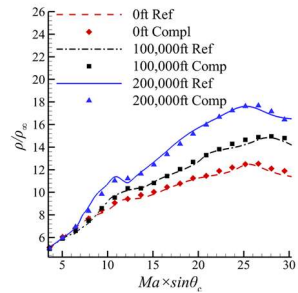
(b) Wall density

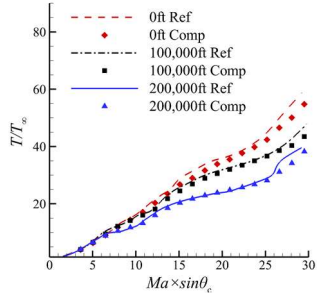

(c) Wall temperature

**Fig. 4**. Comparison of predictions of equilibrium gas (EG) conical flow with reference data by Hudgins [26]

With the conical flow solution verified for an EG, the similitude of the MOC is checked by solving for the curved conical flow in comparison with CFD [20]. The generatrix of the geometry is a cubic curve, the angle between the tangent line of the vertex and the cone axis is 19°, and the end tangent line is parallel to the cone axis. The length of the cone is 10 m, the radius of the bottom surface is 1.6 m, and the other calculation conditions are $H = 70$ km and $Ma = 20$. Fig. 5 shows the predicted pressure contours and the distribution of dimensionless wall pressure $p/p_\infty$ in comparison with the CFD results, and the results validate the MOC for EG flow.

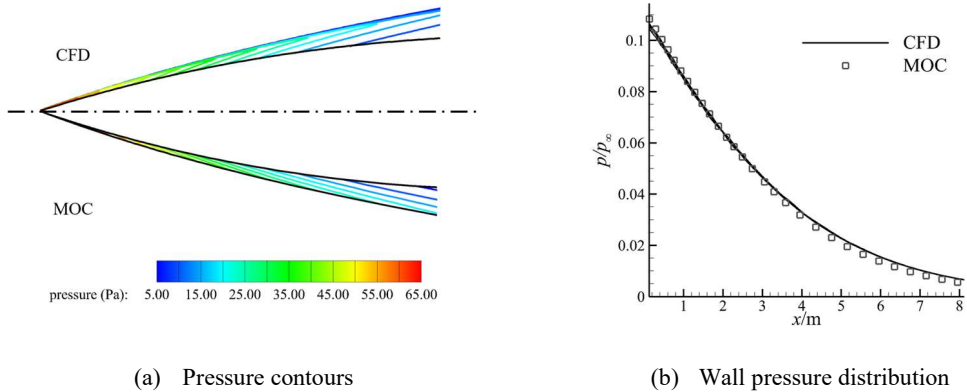

(a) Pressure contours  (b) Wall pressure distribution

**Fig. 5.** Comparison of MOC and CFD results for EG in curved-cone flow field

## 3 Effects of gas model when designing VMOCCW with linear Mach-number distribution

In this section, the VMOCCW design method considering EG effects as discussed in Section 2 is used to generate a waverider. For comparison, a waverider based on the IG model is also generated, allowing investigation of how the different gas models influence the waverider design. Before the primary analysis, the study conducted in [20] is reviewed briefly, which investigated how real-gas effects impact hypersonic waverider design. In [20], the MOC considering EG effects was proposed to generate a series of conical waveriders with different cone half-angles and FCTs, and the influences of the different gas models on the outcomes of the waverider design were studied. The results showed that at high $Ma$, the aerodynamic performances of waveriders designed using the EG and IG models were different, as were the influence degree and pattern of the EG effects on the different waveriders. Specifically, for designs based on straight-cone basic flows, the EG influence increased with the cone half-angle. For instance, when designing with a half-angle of 15°, the relative differences between the two gas models reached 4.05% in volumetric efficiency, 5.28% in volume, and 4.12% in drag coefficient at $Ma = 25$.

While waverider designs considering the EG model were studied in [20], this was done only for constant $Ma$. As a preliminary investigation of real-gas effects in VMW design, this section uses a linear distribution of the design $Ma$ as an example to explore VMOCCW design using different gas models at high altitudes and high $Ma$. For the configurations obtained, the EG model is mainly used for numerical simulations to obtain aerodynamic performances, but for further comparison, simulations are also carried out for the waverider designed using the IG model.

The specific form of the $Ma$ distribution function chosen in this section is indicated by Eq. (1), i.e., a linear increasing distribution from the edge to the symmetric meridian plane with a range of $[Ma_{min}, Ma_{max}] = [6, 20]$. The FCT is the quadratic function $y = Az^2 + B$ with $A = -0.23$ and $B = -0.5425$, and the ICC is the following Bessel function:

$$P(t) = [t^3 \quad t^2 \quad t^1 \quad 1] \begin{bmatrix} -1 & 3 & -3 & 1 \\ 3 & -6 & 3 & 0 \\ 3 & 0 & 0 & 1 \\ 0 & 0 & 0 & 0 \end{bmatrix} \begin{bmatrix} P_0 \\ P_1 \\ P_2 \\ P_3 \end{bmatrix} \tag{18}$$

where $P_0 = (0, -0.5, 0)$, $P_1 = (0, -0.25, -0.26)$, $P_2 = (0, 0.25, -0.26)$, and $P_3 = (0, -0.5, 0)$. The design shock angle is 20°. Based on the above conditions, Fig. 6 shows the VMOCCWs generated using the different gas models. The waverider on the right is designed using the EG model (hereinafter referred to as the EG waverider), while the one on the left is designed using the IG model (hereinafter referred to as the IG waverider). Table 1 gives the geometric characteristics of the two types of waverider, where $\eta = V^{2/3}/S_{wet}$ is the volumetric efficiency, $V$ is the waverider volume, $S_{wet}$ is the surface area, $S_h$ is the horizontal projection area, and $S_b$ is the bottom area. The relative difference is defined as the absolute difference of characteristic $f$ with the two gas models divided by that with the IG model, i.e., $|f_{perf} - f_{eq}|/f_{perf}$, and the same definition applies to subsequent uses of relative difference. The results indicate that the volume and other geometric characteristics of the EG waverider are smaller than those of the IG waverider. This difference arises because when solving for the basic flow, the angle of an oblique shock derived from the EG model is smaller than that from the IG model [24]. As a result, the curved-cone shock for the EG waverider is positioned closer to the basic cone, bringing its leading-edge curve and bottom streamline closer to the basic cone as well. Consequently, the thickness and length of the generated waverider are also reduced. Note that under the linear distribution of the design $Ma$, the volumetric efficiencies of both waveriders are about 0.1, with only a minor difference between them.

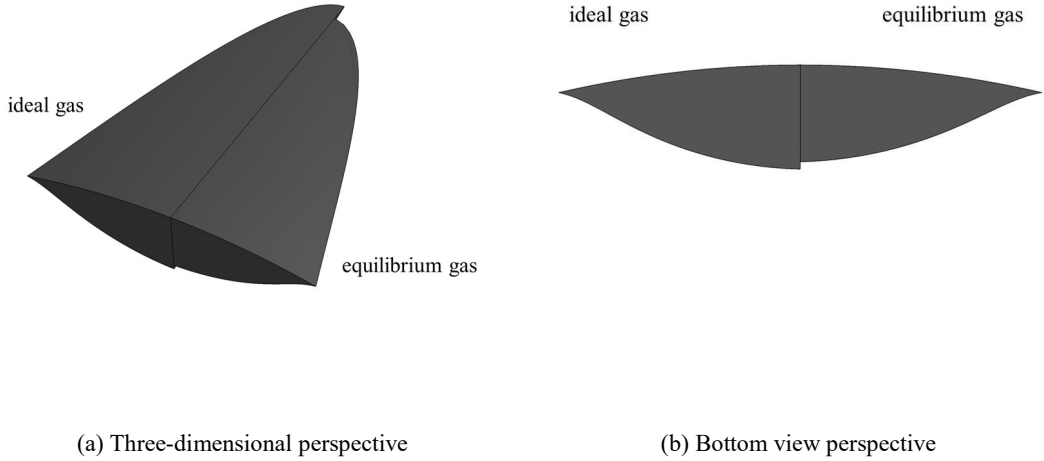

(a) Three-dimensional perspective        (b) Bottom view perspective

**Fig. 6.** VMOCCWs designed with linear *Ma* distribution and different gas models

Table 1 Geometric characteristics of waveriders designed with different gas models, and their relative differences

| Gas model | $V$ [m³] | $S_{wet}$ [m²] | $\eta$ | $S_h$ [m²] | $S_b$ [m²] | Maximum thickness [m] |
|---|---|---|---|---|---|---|
| Ideal | 0.260 | 4.048 | 0.1006 | 1.766 | 0.382 | 0.361 |
| Equilibrium | 0.224 | 3.776 | 0.0977 | 1.658 | 0.352 | 0.336 |
| Relative difference | 13.846% | 6.719% | 2.883% | 6.116% | 7.853% | 6.925% |

To investigate how the choice of gas model affects the aerodynamic characteristics, numerical simulations using the EG model are conducted for the two aforementioned configurations, as well as simulating the IG waverider using the IG model for comparison. The aerodynamic characteristics obtained from the computations include the lift coefficient $C_L$, drag coefficient $C_D$, lift-to-drag ratio $L/D$, absolute pitching moment coefficient $|C_M|$, and reduced pressure center $X_{cp*} = 1 - X_{cp}/L_w$, which denotes the ratio of the distance from the nose to the pressure center at $X_{cp}$ relative to the total length of the waverider $L_w$. The results are shown in Fig. 7, where "Design: EG/IG" indicates design based on the EG or IG model and "Comp: EG/IG" indicates computation based on the EG or IG model, and the results indicate the following. 1) All aerodynamic performances decrease as *Ma* increases. Specifically, $L/D$ becomes relatively stable when *Ma* exceeds 10 (being 4.68 for the EG waverider and 4.58 for the IG waverider); this is because the increases in lift and drag are comparable at high *Ma*. 2) As *Ma* increases, $X_{cp*}$ increases and then decreases, reaching its maximum value around *Ma* = 8. 3) When using real-gas models in computations, differences exist in aerodynamic characteristics and $X_{cp*}$ among waveriders designed using different gas models, with relative differences are given in Table 2. The EG waverider has smaller $C_L$, $C_D$, and $|C_M|$ compared to the IG waverider, while $L/D$ and $X_{cp*}$ are larger for the EG waverider. The largest difference in $|C_M|$ is 13.169%, while the maximum differences in $C_L$ and $C_D$ are 7.027% and 8.877%, respectively. Combining Fig. 7 and Table 2 shows that the relative differences in $C_L$, $C_D$, and $|C_M|$ between the two waveriders increase with *Ma*, while the relative differences in $L/D$ and $X_{cp*}$ change little with increasing *Ma*. 4) When the different gas models are used to compute the aerodynamic characteristics and $X_{cp*}$ for the IG waverider, the results are essentially consistent. This suggests that the differences in simulations by various gas models for the IG waverider should not be used to assess the differences between gas models in the design process. Overall, under a linear distribution of the design $Ma$, the EG waverider has smaller $C_L$ and $C_D$ compared to the IG waverider, but its $L/D$ is higher and its $X_{cp*}$ is located further aft. In general, the EG waverider exhibits better flight stability.

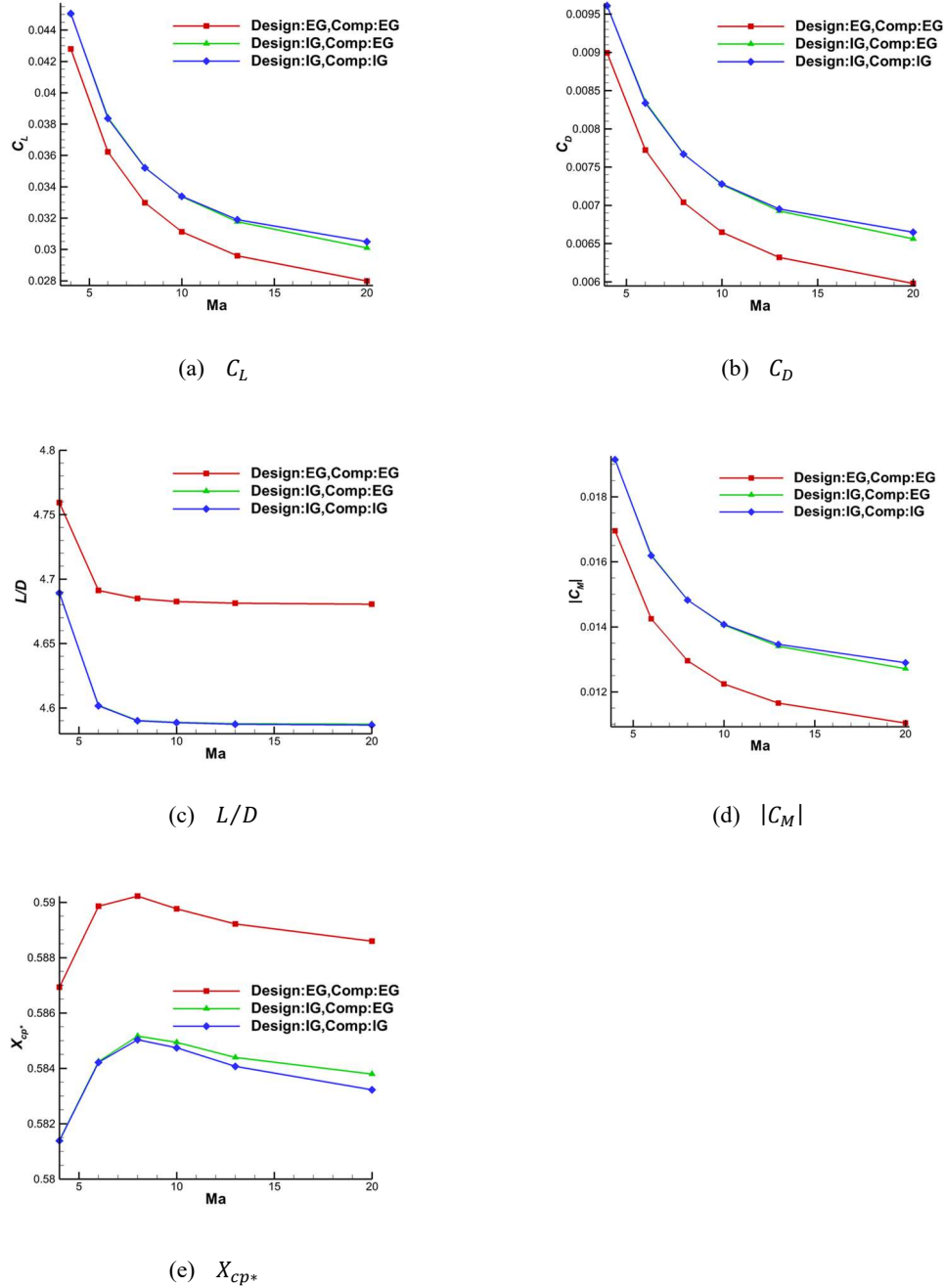

(a) $C_L$

(b) $C_D$

(c) $L/D$

(d) $|C_M|$

(e) $X_{cp*}$

**Fig. 7.** Distributions of aerodynamic characteristics of VMOCCW designed by different gas models

Table 2 Relative differences in aerodynamic characteristics of waveriders designed with different gas models using EG simulations

| $Ma$ | 4 | 6 | 8 | 10 | 13 | 20 |
|---|---|---|---|---|---|---|
| $C_L$ | 5.009% | 5.748% | 6.390% | 6.7161% | 6.871% | 7.027% |
| $C_D$ | 6.409% | 7.541% | 8.281% | 8.584% | 8.734% | 8.877% |
| $L/D$ | 1.496% | 1.939% | 2.062% | 2.043% | 2.041% | 2.031% |
| $|C_M|$ | 11.436% | 12.149% | 12.633% | 12.887% | 13.032% | 13.169% |
| $X_{cp*}$ | 0.951% | 0.960% | 0.864% | 0.824% | 0.826% | 0.822% |

Fig. 8 shows the lower-surface pressure distributions and cross-section contours of the waveriders designed with the different gas models at $Ma$ = 4–20 and $H$ = 70 km. For $Ma \leq 8$, pressure overflow occurs at the edges of the VMWs. Notably, at $Ma = 8$, the EG waverider has no pressure overflow at the edges, while the IG waverider still exhibits pressure overflow at the edges. This suggests that at $Ma = 8$, the EG waverider has a better $L/D$ compared to the IG waverider, which is consistent with the results shown in Fig. 7(c). As $Ma$ increases further, neither waverider type shows pressure overflow at the edges. In addition, Fig. 8 shows that there are some fluctuations in the numerical results, although the numerical method used herein employs the minmod TVD limiter. Note that similar fluctuations were also observed in CFD studies such as [27], which indicates a common issue that must be improved in subsequent studies.

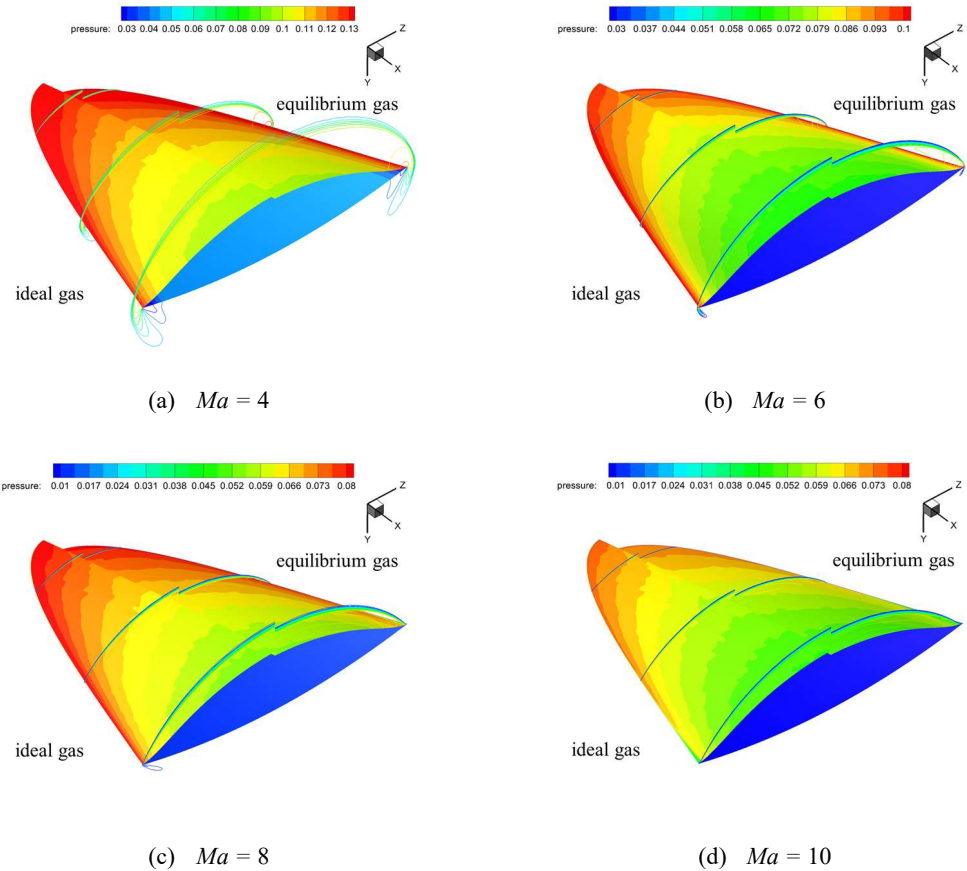

(a)  $Ma = 4$     (b)  $Ma = 6$

(c)  $Ma = 8$     (d)  $Ma = 10$

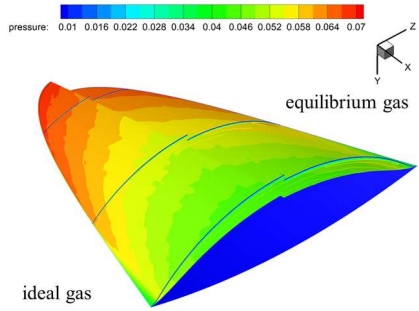 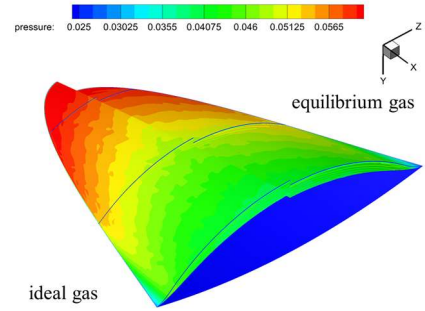

(e)  *Ma* = 13　　　　　　　　　　　　　　(f)  *Ma* = 20

**Fig. 8.** Pressure contours on lower surface and those on cross-sections of waveriders by different gas models from computations using EG model

To show further the quantitative differences in pressure distribution of the waveriders designed with the different gas models, Fig. 9 shows the distribution of the pressure coefficient $C_p$ on the bottom cross-section ($x = 0$) of the aforementioned two waverider types at *Ma* = 4, 6, 8, and 13. As can be seen, the distribution of $C_p$ on the bottom cross-section of the waveriders designed using the two gas models follows a similar pattern. Specifically, for both waverider shapes at different *Ma*, the pressure coefficient changes smoothly near the symmetry plane, while the distribution generally increases toward the edges. This trend is consistent with the pressure contours in Fig. 8. Additionally, it is observed that at all *Ma*, $C_p$ drops suddenly near the edges of the waveriders. This is due to grid singularities at the waverider edges, a common phenomenon in CFD for sharp leading edges. Notably, when *Ma* = 8 and especially 13, the $C_p$ distribution near the edge show a maximum followed by a limited decrease, which differs from the sharp drop observed at *Ma* = 4 and 6.

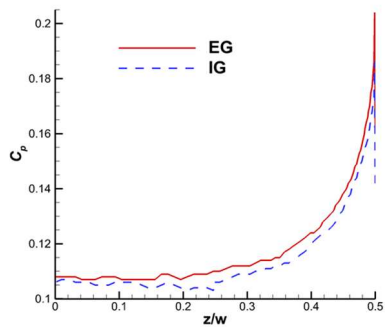 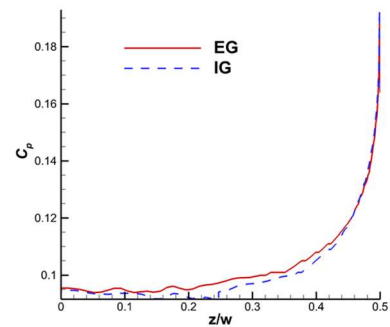

(a)  *Ma* = 4　　　　　　　　　　　　　　(b)  *Ma* = 6

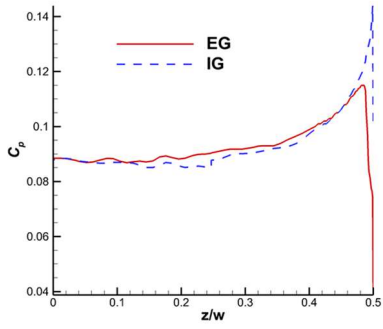
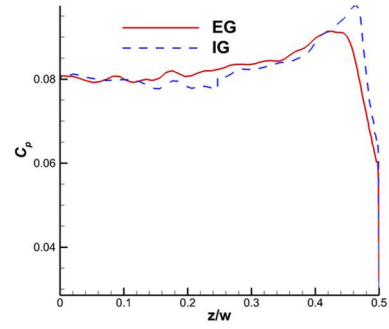

(c) $Ma = 8$  (d) $Ma = 13$

**Fig. 9.** $C_p$ along lower bottom surface of waveriders designed with different gas models from computations using EG model

Fig. 10 shows contours of the dimensionless density $\rho/\rho_\infty$ on the lower surface of the IG waverider at the chosen conditions of $H = 70$ km and $Ma = 13$ and 20, where computations are performed using the two gas models. Moreover, Fig. 11 shows the distributions of specific heat ratio $\tilde{\gamma}$ on the lower surface by computations using the EG model under the same conditions. These show again that real-gas effects alter physical quantities such as surface density and $\gamma$ at higher $Ma$. Specifically, given $\gamma = 1.4$ in IG flow, the effective $\tilde{\gamma}$ of EG flow varies from 1.36 to 1.40.

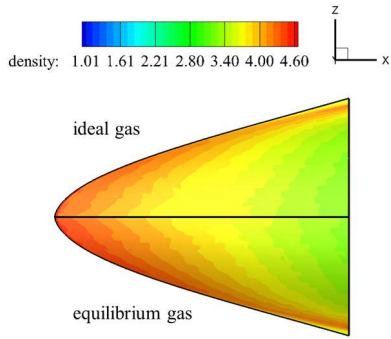
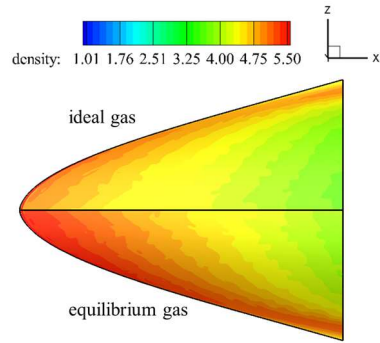

(a) $Ma = 13$  (b) $Ma = 20$

**Fig. 10.** Density contours on lower surface of IG waverider under conditions of both gas models

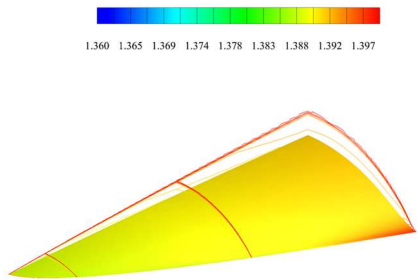
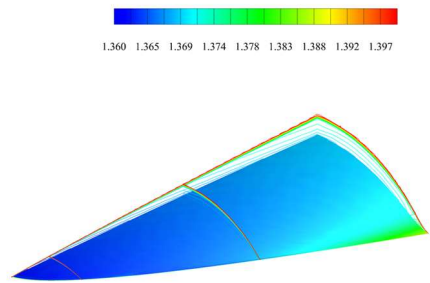

(a) $Ma = 13$  (b) $Ma = 20$

**Fig. 11.** Contours of $\tilde{\gamma}$ on lower surface of IG waverider in computations of EG flow

# 4 Rational function for *Ma* distribution and comparable functions

In this section, we propose an RF for the *Ma* distribution with the EG model, and we compare it with other methods. Before that, we briefly review the study in [17] of *Ma* distribution methods, where the impacts of several methods were discussed.

## 4.1 Review of investigation of effects of some *Ma* distributions [17]

Because of the various *Ma* distributions used in VMW design, some studies have been conducted on how the choice of distribution affects the waverider design. Below is a brief review of related research reported in [17], which studied the effects of some distributions on the configuration and aerodynamic performance of an osculating-cone VMW. The study was carried out under IG flows and examined the following three aspects of the distribution function: linearity/nonlinearity, monotonicity, and concavity/convexity. The chosen functions were linear functions, sine functions, and cosine functions for *Ma* in a given range, and these were used to design and compare the corresponding geometric configurations. The results showed that the distributions had different impacts on the waverider characteristics. In detail, the impacts on the geometric configuration were as follows. 1) Under the same monotonicity, the waverider designed with a linear distribution enclosed that designed with a nonlinear concave function (e.g., $1 - \cos x$ or $1 - \sin x$), while it was contained by the waverider designed with a convex function (e.g., $\sin x$ or $\cos x$). 2) Under the same conditions, the waverider designed with a monotonically increasing function was longer and thicker in the middle of the configuration and narrower at the edge of the configuration than the one designed with a monotonically decreasing function. 3) Under the same conditions, the waverider designed with a convex function contained that designed with a concave function and had larger $\eta$. The impacts on aerodynamic performance were as follows. 1) Under the same monotonicity, the waverider designed with a linear distribution had larger $L/D$ than that of the one designed with a convex function but smaller than that of the one designed with a concave function. 2) The waverider designed with a monotonically decreasing distribution had higher $L/D$ than the one designed with a monotonically increasing distribution. The former always had lower $C_L$ at small angle of attack, while the latter had higher $C_L$ at large angle of attack. Additionally, the former always exhibited lower $C_D$.

Although [17] made progress in studying *Ma* distributions, the diversity of distribution functions means that further research is needed to optimize the function in order to achieve a waverider with balanced high $L/D$ and $\eta$ concurrently. In addition, real-gas effects on waverider design are also worth studying in conjunction with the distribution method. These issues are discussed in detail in Section 4.2.

## 4.2 Designing a waverider using a rational function

To achieve a VMW with balanced $\eta$ and $L/D$ considering the *Ma* distributions in the design as discussed in Section 4.1, we propose the following parameterized RF:

$$Ma = \frac{Ma_{max}-Ma_{min}}{c_2+\frac{(c_2-0.5)^2}{1.25-c_2+c_1(0.5-c_2)0.5^m}}\left(c_2 + \frac{(z^*-c_2+0.5)^2}{(z^*-c_2+0.5)+c_1\ (z^*-c_2+0.5)\ (z^*+0.5)^m+3\ (z^*+0.5)^2}\right) + Ma_{min} \qquad (19)$$

where the independent variable is $z^* = z/w \in [-0.5, 0]$, $w$ is the spanwise length of the waverider at the bottom, and $z^* = -0.5$ or $0$ corresponds to the edge or the symmetry plane of the waverider, respectively. In Eq. (19), $Ma_{min}$ and $Ma_{max}$ are the design inputs for defining the $Ma$ range, and $c_1$, $c_2$, and $m$ are the control parameters for adjusting the distribution of the function. This function is initially applied in the WENO scheme based on the mapping function [28], where the associated parameters adjust the flatness of the function at the interval endpoints of the independent variable and the degree of transition of the function within the interval. Specifically, in the present design, $c_1$ affects the degree of flatness of the function at the two endpoints, $c_2$ influences the transition location of $Ma$ from the symmetry plane to the edge, and $m$ controls the slope of the transition. By adjusting the control parameters, the distribution function achieves a relatively flat variation near the two endpoints of the design $Ma$. Additionally, it is concave near the edge and convex near the symmetry plane, aiming to leverage the respective advantages of concave and convex functions. In this study, the input range of $Ma$ is $[Ma_{min}, Ma_{max}] = [6, 20]$; Additionally, to compare the differences in the waveriders designed with RFs with different control parameters, two sets of parameter are chosen, i.e., $\{c_1, c_2, m\} = \{5e4, 14.05, 7.5\}$ with RF1 and $\{c_1, c_2, m\} = \{5e6, 14, 11\}$ with RF2. The function distributions are shown in Fig. 12, and compared to RF1, RF2 is flatter near the two endpoints and undergoes a steeper transition in the turning region.

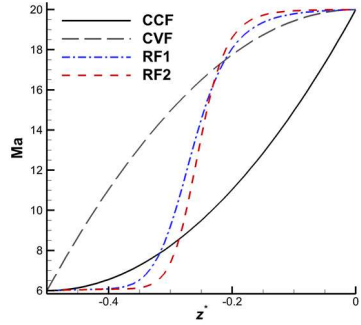

**Fig. 12.** Various $Ma$ distributions in $z^*$ for VMOCCW design

To compare the proposed rational distribution with typical ones such as concave and convex functions, the following monotonically increasing quadratic concave function (CCF) and convex function (CVF) are chosen as references as plotted in Fig. 12:

$$CCF: Ma = Ma_{min} + (Ma_{max} - Ma_{min}) \cdot \frac{(z^* + 0.5)^2}{0.25} \tag{20}$$

$$CVF: Ma = Ma_{max} + (Ma_{min} - Ma_{max}) \cdot \frac{z^{*2}}{0.25} \tag{21}$$

In the present design, the forms of the FCT and ICC and the design shock angle are the same as those in Section 3. The waveriders designed using the four distributions (see Fig. 12) are shown in Fig. 13, for which the IG and EG models were used. Fig. 13 shows that under the same conditions, the length and thickness of the IG waverider are greater than those of the EG waverider, implying that the former has a larger volume and surface area than the latter. Fig. 13(a) shows that the leading-edge curves of the RF1 and RF2 waveriders expand rapidly in the spanwise direction initially and then stagnate around $x = -600$ mm, whereupon the curves expand rapidly again until the bottom position. Thus, the corresponding waveriders form a wing-like structure that is scarce in current waverider designs. Fig. 13(a) and (b) show that the leading-edge and trailing-edge curves of the RF1 and RF2 waveriders coincide with those of the CVF and CCF ones at the symmetric meridian and outer edge, which indicates that when the same $Ma$ is used to define the osculating plane, the streamlines and corresponding waverider configurations are identical. Fig. 13(c) shows the cross-section contours of the waveriders at $x = -500$ mm, which are consistent with those in Fig. 13(b). Fig. 13(d) shows that the contours at $x = -850$ mm on the lower surface of the RF1 and RF2 waveriders and the CVF waverider basically coincide, which indicates that the wing-like structure of the RF1 and RF2 waveriders degenerates into the traditional configuration near the head of the waverider. In addition, when using the same gas model in the design, the CVF waverider encloses the CCF waverider, which is as concluded in [17].

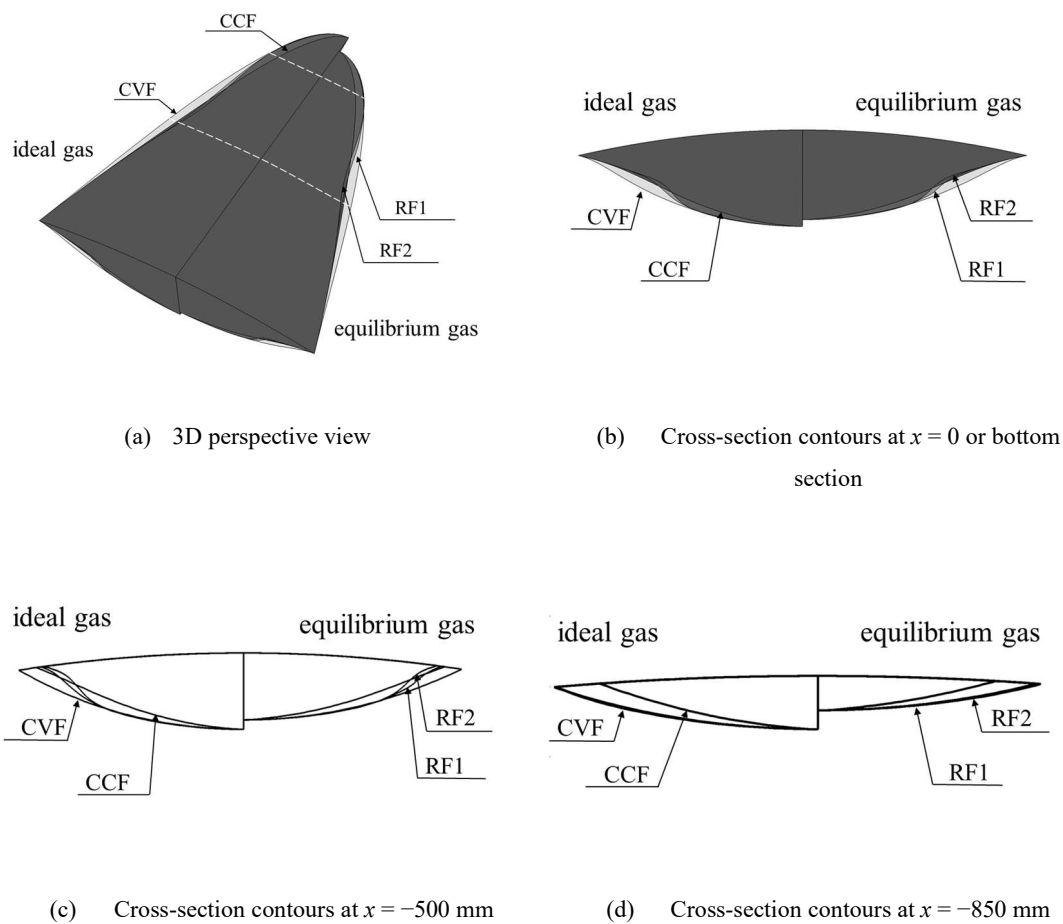

(a)  3D perspective view  
(b)  Cross-section contours at $x = 0$ or bottom section  
(c)  Cross-section contours at $x = -500$ mm  
(d)  Cross-section contours at $x = -850$ mm

**Fig. 13.** Configurations of CCF, CVF, RF1, and RF2 waveriders designed with different gas models, and their cross-section contours at different locations as indicated by the dotted lines in (a)

Figs. 12 and 13(b) indicate that the local design $Ma$ of the CVF, RF1, and RF2 waveriders is larger than that of the CCF waverider when the spanwise location is near the symmetric meridian plane, and thus the former waveriders appear thicker there. The local design $Ma$ of the CCF, RF1, and RF2 waveriders near the edge is smaller than that of the CVF waverider, and the former waverider configurations are thinner. This is because in the same osculating plane when the shock angle is constant, a higher design $Ma$ results in streamlines that are farther from the object plane, leading to a thicker waverider configuration. In general, the symmetric meridian plane is the thickest part of the whole geometry, accounting for a large proportion of the volume space. Therefore, in designs using the same gas model, the volumes of the RF1 and RF2 waveriders are between those of the CCF and CVF waveriders. Table 3 gives the geometric characteristics of the waveriders designed using the different distribution functions and gas models, and the results are quantitatively consistent with the above discussion. Moreover, Table 3 shows that the $\eta$ values of the RF1 and RF2 waveriders are improved compared with those of the CCF and CVF waveriders.

Table 3 Geometric characteristics of waveriders designed using different $Ma$ distribution functions and gas models

| Gas model | Waverider | $V$ [m$^3$] | $S_{wet}$ [m$^2$] | $\eta$ | $S_h$ [m$^2$] | $S_b$ [m$^2$] |
|---|---|---|---|---|---|---|
| Ideal | CCF | 0.239 | 3.847 | 0.1000 | 1.678 | 0.362 |
| | CVF | 0.270 | 4.122 | 0.1013 | 1.795 | 0.395 |
| | RF1 | 0.256 | 3.964 | 0.1017 | 1.717 | 0.376 |
| | RF2 | 0.253 | 3.942 | 0.1014 | 1.703 | 0.372 |
| Equilibrium | CCF | 0.208 | 3.628 | 0.0968 | 1.595 | 0.334 |
| | CVF | 0.234 | 3.871 | 0.0982 | 1.697 | 0.363 |
| | RF1 | 0.225 | 3.737 | 0.0986 | 1.632 | 0.347 |
| | RF2 | 0.221 | 3.713 | 0.0983 | 1.620 | 0.343 |

From an engineering perspective, the effective volume of the waverider should be concentrated on the symmetric plane as much as possible, which yields a larger and more-efficient loading space for the aircraft. Taking the section at the waverider bottom as an example and considering the section area $S(z^*)$ distributed with $z^*$, the ratio of $S(z^*)$ to the whole area $S_b$ is assessed to determine where the ratio is close to unity, e.g., 0.8. Obviously, the smaller the value of $z^*$ for that location, the higher the concentration proportion near the symmetric plane. Based on the above understanding, Fig. 14 shows $S(z^*)/S_b$ for the RF2, CCF, and CVF waveriders as calculated using the EG model. The aforementioned values of $z^*$ are determined by where the line of constant ratio 0.8 intersects the $S(z^*)/S_b$ distributions of the waveriders. The results show that $z^*$ of the RF2 waverider is the smallest, indicating that its bottom area is more concentrated near the symmetric plane and therefore better for a larger loading space. In turn, this waverider naturally has a smaller area or thickness near the edge; this makes the waverider flatter there, and it has a wing-like structure accordingly.

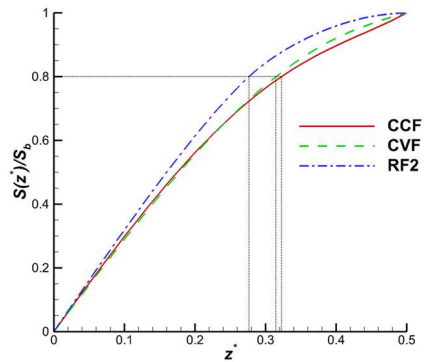

**Fig. 14.** Area ratios at bottom section of waveriders designed with EG model and different $Ma$ distributions

To investigate the aerodynamic characteristics such as $L/D$ of the VMW designed with the

RF $Ma$ distribution, CFD simulations are conducted using the EG model, and the results are shown in Fig. 15. To analyze how the choice of gas model affects the aerodynamic characteristics, results for the EG and IG waveriders are also provided for comparison. For waveriders designed with different gas models but the same $Ma$ distribution, their aerodynamic characteristics show the following. 1) Using both gas models in the design, the characteristics of the waveriders with the four $Ma$ distributions decrease with increasing $Ma$, and $L/D$ tends to change slowly when $Ma$ exceeds 10, being 4.65–4.75 for the EG waveriders and 4.51–4.60 for the IG waveriders. 2) With increasing $Ma$, $X_{cp*}$ increases and then decreases, reaching a maximum value around $Ma$ = 6 for the CCF, RF1, and RF2 waveriders and around $Ma$ = 8 for the CVF waverider. 3) When different gas models are used in the design, the aerodynamic characteristics with the four $Ma$ distributions are different, with $C_L$, $C_D$, and $|C_M|$ of the EG waveriders being smaller than those of the IG waveriders, but $L/D$ being larger. For waveriders designed with the same gas model but different $Ma$ distributions, the CFD results indicate the following. 1) $C_L$, $C_D$, and $|C_M|$ follow the order of CVF > RF1 > RF2 > CCF. 2) $L/D$ and $X_{cp*}$ follow the order of CCF > RF2 > RF1 > CVF. 3) Combining these findings with the geometric characteristics in Table 3 reveals the following trade-off relationship between volume and $L/D$: the CCF waverider has the highest $L/D$ and the smallest volume, while the CVF waverider has the largest volume and the smallest $L/D$; $L/D$ and the volumes of the RF1 and RF2 waveriders are between those of the other two but with better $\eta$. In addition, the RF1 waverider has larger $\eta$ but smaller $L/D$ than those of the RF2 waverider. Further optimization of the RF function is expected to give a waverider with an even better $L/D$ and $\eta$. Overall, the RF1 and RF2 waveriders exhibit more-balanced aerodynamic performance and geometric characteristics.

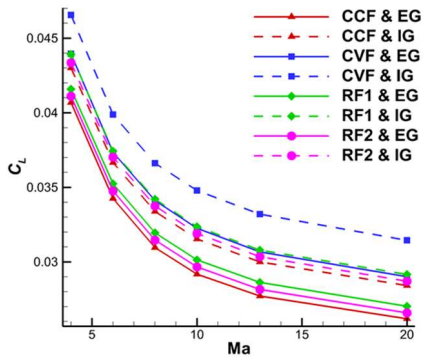

(a) $C_L$

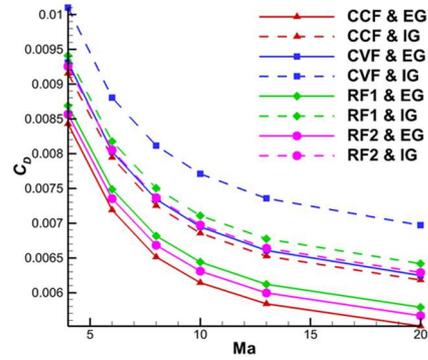

(b) $C_D$

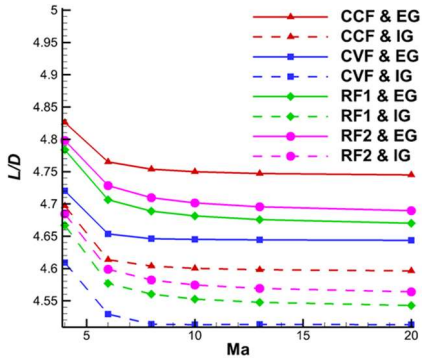

(c) $L/D$

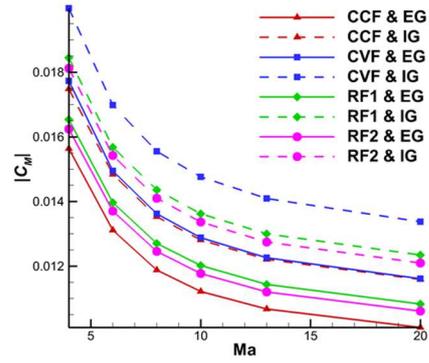

(d) $|C_M|$

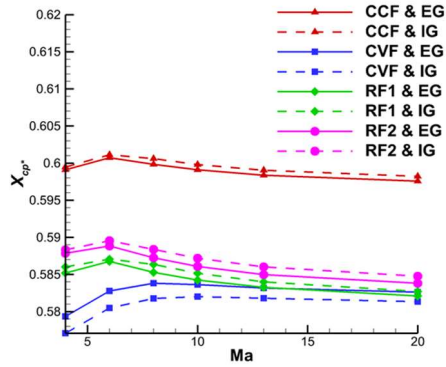

(e) $X_{cp*}$

**Fig. 15.** Comparison of aerodynamic characteristics of CCF, CVF, RF1, and RF2 waveriders designed with different gas models

To further demonstrate the impact of different gas models in design, the CVF, CCF, and RF2 waveriders designed using the different gas models are used to calculate the relative differences ($|f_{perf} - f_{eq}|/f_{perf}$) of aerodynamic characteristics as given in Table 4. As can be seen, the waveriders designed with the same $Ma$ distribution but different gas models exhibit differences in $C_L$, $C_D$, $L/D$, and $|C_M|$ in real-gas flow. These differences increase with $Ma$, whereas that in $L/D$ does not change much. In addition, the magnitudes of the relative difference in $X_{cp*}$ are small. Specifically, the respective maximum differences of $C_L$, $C_D$, and $|C_M|$ due to the choice of gas model are 7.863%, 10.742%, and 12.805% for the CCF waverider, 7.791%, 10.383%, and 13.189% for the CVF waverider, and 7.404%, 9.888%, and 12.311% for the RF2 waverider. These results further demonstrate that the choice of gas model in the design can significantly affect the waverider outcomes.

Table 4 Relative differences of aerodynamic characteristics of CCF, CVF, and RF2 waveriders designed with different gas models

| Waverider | $Ma$ | 4 | 6 | 8 | 10 | 13 | 20 |
|---|---|---|---|---|---|---|---|
| CCF | $C_L$ | 5.397% | 6.566% | 7.252% | 7.471% | 7.666% | 7.863 % |
|  | $C_D$ | 7.932% | 9.533% | 10.176% | 10.384% | 10.563% | 10.742% |
|  | $L/D$ | 2.753% | 3.280% | 3.256% | 3.250% | 3.240% | 3.226% |
|  | $|C_M|$ | 10.539% | 11.633% | 12.201% | 12.420% | 12.613% | 12.805% |
|  | $X_{cp*}$ | 0.058% | 0.066% | 0.127% | 0.119% | 0.113% | 0.109% |
| CVF | $C_L$ | 5.594% | 6.340% | 6.844% | 7.258% | 7.587% | 7.791% |
|  | $C_D$ | 7.822% | 8.839% | 9.497% | 9.896% | 10.187% | 10.383% |
|  | $L/D$ | 2.417% | 2.742% | 2.932% | 2.927% | 2.896% | 2.892% |
|  | $|C_M|$ | 11.275% | 11.993% | 12.450% | 12.734% | 12.981% | 13.189% |
|  | $X_{cp*}$ | 0.388% | 0.397% | 0.350% | 0.280% | 0.233% | 0.222% |
| RF2 | $C_L$ | 5.201% | 6.121% | 6.768% | 7.001% | 7.201% | 7.404% |
|  | $C_D$ | 7.450% | 8.686% | 9.290% | 9.514% | 9.703% | 9.888% |
|  | $L/D$ | 2.430% | 2.809% | 2.780% | 2.777% | 2.771% | 2.756% |
|  | $|C_M|$ | 10.330% | 11.151% | 11.666% | 11.893% | 12.100% | 12.311% |
|  | $X_{cp*}$ | 0.079% | 0.117% | 0.195% | 0.191% | 0.178% | 0.164% |

To further compare and analyze the differences in aerodynamic performance of waveriders designed with the RF *Ma* distribution and other methods, Fig. 16 shows the pressure contours on the lower surface and its front-to-end view with sectional contours shown at the bottom of the CCF, CVF, and RF2 waveriders at *Ma* = 6, 13, and 20. To analyze how the choice of gas model influences the aerodynamic performance, Fig. 16 also provides a comparison regarding waveriders designed using the two gas models. Fig. 16 shows the following. 1) For the waveriders designed using the same gas model and *Ma* distribution, the bottom pressure spill-out decreases gradually with increasing *Ma* because of the stronger compressibility at high *Ma* and the greater proximity of the shock wave to the wall. 2) For the waveriders designed using different gas models but the same *Ma* distribution, the following characteristics are observed: the pressure spill-out at the bottom of the EG waverider is smaller than that of the IG waverider (e.g., see the case for *Ma* = 6), which indicates a larger $L/D$ of the former and is consistent with Fig. 15(c). 3) For the waveriders designed using the same gas model but different *Ma* distributions, taking the example of *Ma* = 6, the high-pressure region of the CCF and CVF waveriders is located in the head and the entire leading edge of the vehicle, while for the RF2 waverider, the high-pressure region is located mainly in the head and at the position of the wing-like structure, with some low-pressure distribution in part of the leading edge. This is because the leading edge of the RF2 waverider extends in the spanwise direction to form a wing-like structure, and the rapid thinning of the configuration results in expansion waves. Compared with the CVF waverider, the CCF and RF2 waveriders have smaller bottom pressure spill-out, which indicates that the latter have larger $L/D$ as shown in Fig. 15(c). The likely reason for this is that the concave configuration on the lower surface provides a larger buffer space to accommodate the shock wave. With increasing *Ma* (*Ma* = 13 or 20), the high-pressure region of the three waveriders is located mainly in the head, and the area decreases. The high-pressure region of the RF2 waverider also decreases at the root of the wing-like structure with increasing *Ma*. Note that when *Ma* is high, the RF2 waverider has a region with rather high pressure on the inner side of the wing-like structure; this could be a new lift mechanism for this type of waverider, and it would be worth investigating how to utilize and optimize this mechanism for increased lift and $L/D$.

|     | CCF | CVF | RF2 |
|-----|-----|-----|-----|
| Ma6 | 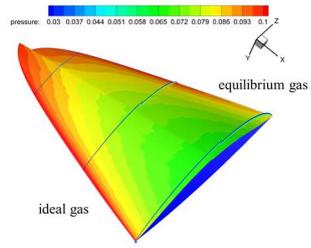<br>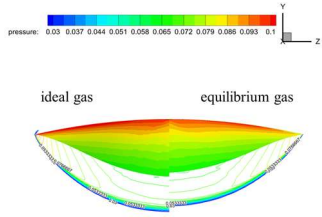 | 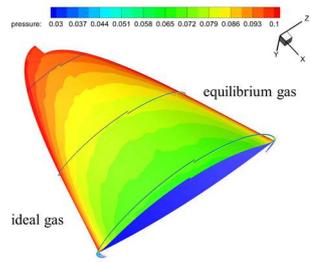<br>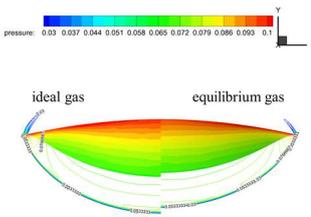 | 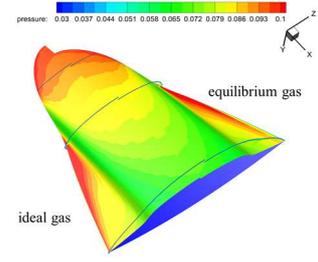<br>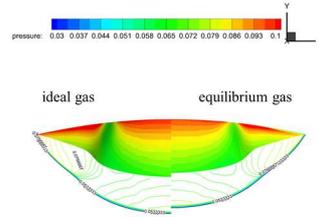 |
| Ma13 | 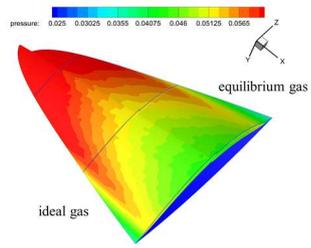<br>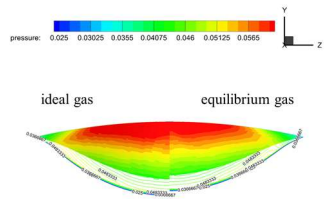 | 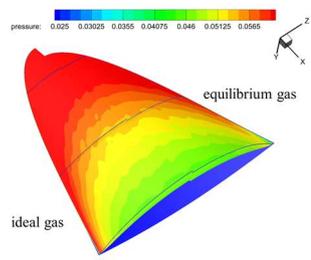<br>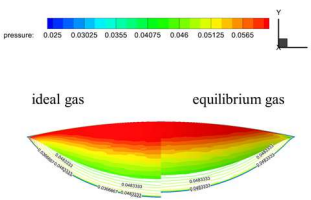 | 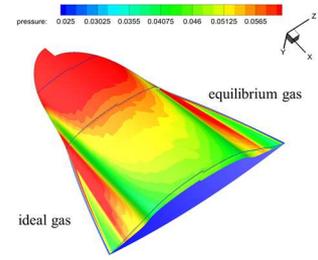<br>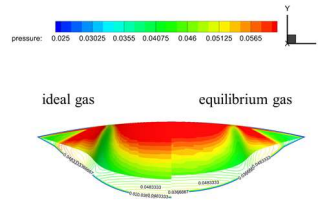 |
| Ma20 | 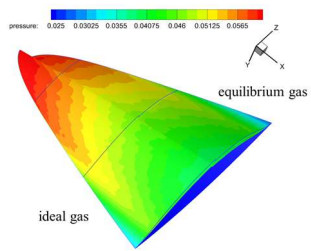<br>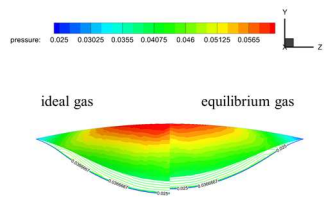 | 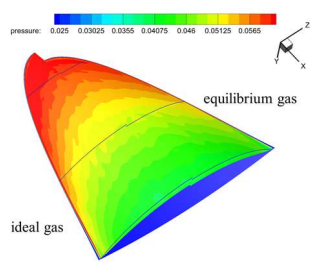<br>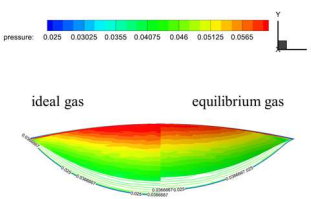 | 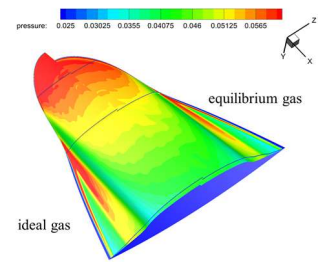<br>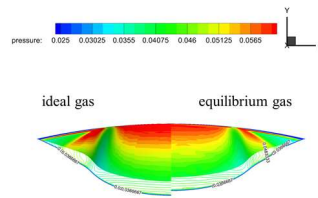 |

Fig. 16. Pressure contours on lower surface and its front-to-end view with sectional contours shown at bottom at different $Ma$ for CCF, CVF, and RF2 waveriders designed with different gas models

To show quantitatively the differences in the pressure distributions of waveriders designed with different $Ma$ distributions, Fig. 17 shows the distributions of the pressure coefficient on the lower surface at various cross-sections ($x = 0, -300$ mm, $-500$ mm) of the CCF, CVF, and RF2 waveriders at $Ma$=13. At the same time, to show the effects of the choice of gas model, the distributions of the waveriders designed with both gas models are also shown. Fig. 17 also indicates that $C_p$ drops suddenly near the edges, which is likely caused by geometric singularities at the edges in the CFD calculation. Also, for the waveriders designed with the same gas model but different $Ma$ distributions, the $C_p$ distributions at the cross-sections near the symmetric meridian are generally consistent, with relatively flat profiles. However, noticeable differences occur from $z/w = 0.1$ to the edges. Specifically, $C_p$ of the CCF waverider increases and then decreases, the CVF waverider has a monotonically increasing distribution, and the RF2 waverider has a distribution that decreases and then increases following a local minimum [e.g., see Fig. 17(c)]. Furthermore, at the sections for $x = 0$ and $-300$ mm, the first and second local minima occur, which are separated by a local maximum [see Fig. 17(a) and (b)]. Considering the pressure contours on the lower surface in Fig. 16, it is clear that the first local minimum is due to the expansions engendered by the wing-like structure, which reduce the pressure. The local maximum at the $x = 0$ and $-300$ mm sections corresponds to the high-pressure region at the inner side of the wing-like structure of the RF2 waverider. The second local minimum occurs near the edge of the wing-like structure, and it is sandwiched between the aforementioned maximum and the outer edge with high pressure. For the designs with different gas models but the same $Ma$ distribution, the corresponding sectional $C_p$ indicates similar distributions.

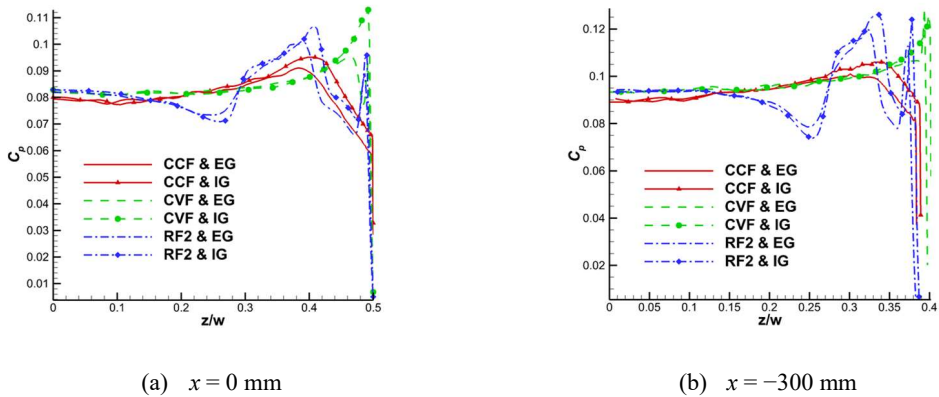

(a) $x = 0$ mm         (b) $x = -300$ mm

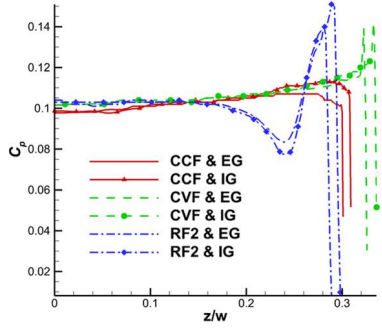

(c)　$x = -500$ mm

**Fig. 17.** $C_p$ distributions on lower surface at three cross-sections of CCF, CVF, and RF2 waveriders designed with different gas models at $Ma = 13$

In short, the RF1 and RF2 waveriders have wing-like characteristics and more-balanced geometric and aerodynamic characteristics than the CCF and CVF waveriders. Also, disparities of characteristics exist between the RF1 and RF2 waveriders, which indicate that the layout can be further optimized by adjusting the control parameters.

### 4.3 Aerodynamic performances of waveriders under off-design conditions

To explore the performances of wide-speed-range waveriders at off-design $Ma$, the CCF, CVF, and RF1 EG waveriders are selected in this subsection for CFD simulations at $Ma = 4$. Because this $Ma$ is relatively low and so the real-gas effects are weak, the IG model is used in all the computations. The aerodynamic characteristics are given in Table 5, and as can be seen, the values of $C_L$, $C_D$, $L/D$, $|C_M|$, and $X_{cp*}$ for the RF1 waverider are between those for the CCF and CVF waveriders.

Table 5 Aerodynamic characteristics of CCF, CVF, and RF1 EG waveriders at $Ma = 4$

| Waverider | $C_L$ | $C_D$ | $L/D$ | $|C_M|$ | $X_{cp*}$ |
|---|---|---|---|---|---|
| CCF | 0.0407 | 0.00843 | 4.828 | 0.0156 | 0.595 |
| CVF | 0.0440 | 0.00932 | 4.721 | 0.0178 | 0.575 |
| RF1 | 0.0416 | 0.00869 | 4.787 | 0.0165 | 0.581 |

Fig. 18 shows the pressure contours on the lower surface and its front-to-end view superimposed with the bottom-section contours. Because of the difference between the range of the design $Ma$ at the edge and the chosen off-design $Ma$, the three waveriders have relatively serious overflows. The pressure contours on the lower surface show that the high-pressure region of the CCF waverider is located mainly on the narrow head and the entire leading edge [see Fig. 18(a)]. The performance of the CVF waverider is similar to that of the CCF waverider; however, the high-pressure region in the wider head of the waverider appears larger, but that at the outer edge near the bottom is smaller [see Fig. 18(b)]. Because of the wing-like structure that appears in the RF1 waverider, the expansions that occur there cause a low-pressure region near the leading edge. Also, the high-pressure region in the wing-like part indicates a larger area [see Fig. 18(c)]. Although exhibiting noticeable overflow, this configuration can still maintain a relatively high $L/D$. In general, under off-design conditions, the aerodynamic performance of the RF1 waverider is still between those of the CCF and CVF waveriders.

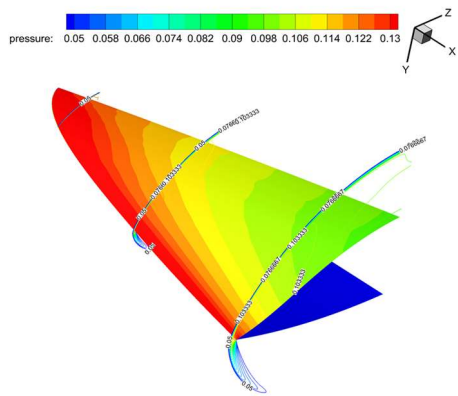 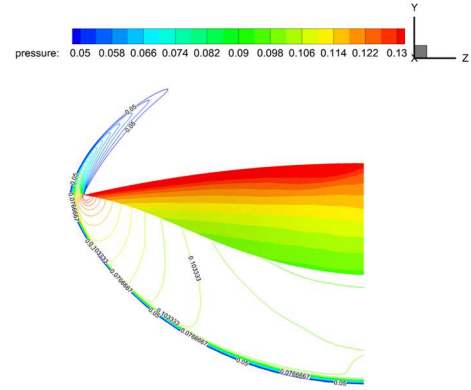

(a) CCF waverider

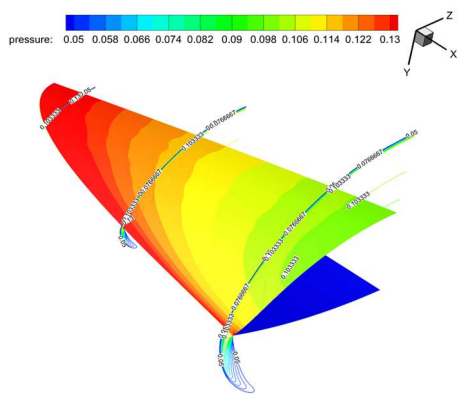 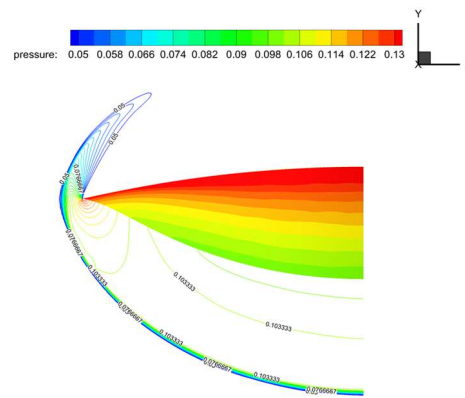

(b) CVF waverider

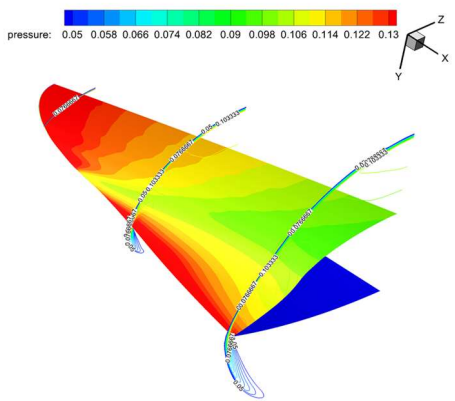 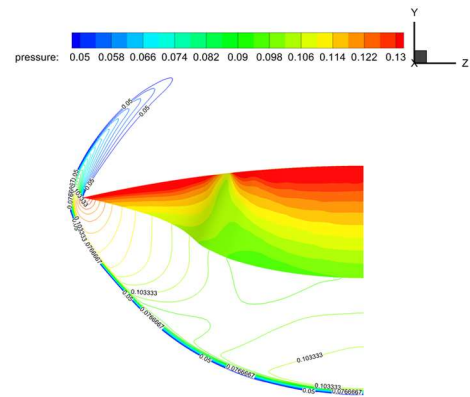

(c) RF1 waverider

**Fig. 18.** Pressure contours on lower surface as well as at three sections (left) and those at bottom section together with front-to-end view of surface contours (right) of CCF, CVF, and RF1 EG waveriders at $Ma = 4$

# 5 Conclusions

Concerning real-gas effects and the requirement for a wide speed range in the design of hypersonic waveriders, reported herein were research and analysis of the design of a VMOCCW considering EG effects in the MOC based on [20]. First, taking a linear *Ma* distribution as an example, the influences of different gas models on the design were investigated. Next, a parameter-adjustable RF was proposed, with quadratic concave and convex increasing functions chosen for comparison, using which VMOCCWs were designed. In this way, the influence of the *Ma* distribution on the layout and aerodynamic characteristics was further studied. Finally, the aerodynamic characteristics at off-design *Ma* were discussed. We draw the following conclusions.

(1) When different gas models are used in VMOCCW design, the obtained waveriders have different geometric characteristics. Specifically, the length, area, and volume of EG waveriders are smaller than those of IG waveriders; the volume difference between the two different types of waverider is relatively large, reaching 13.846% in the case of a linear distribution. Also, differences occur in the aerodynamic properties; $C_L$, $C_D$, and $|C_M|$ of EG waveriders are lower than those of IG waveriders, and the discrepancy increases with *Ma*. However, $L/D$ of EG waveriders is greater than that of IG waveriders. In waveriders designed with a linear *Ma* distribution, $|C_M|$ with the two gas models differs by a maximum of 13.169%. Therefore, real-gas effects are considered significant in hypersonic waverider design.

(2) A parametric RF for the *Ma* distribution was proposed for VMW design. The RF has adjustable flatness at the symmetric meridian plane and edge, and the position and degree of its transition from the symmetric meridian plane to the edge are controllable. Using the RF, two waveriders were generated using RF1 and RF2 with different control parameters for the given *Ma* range of [6, 20]. In addition to the basic waverider characteristics, the two waveriders had a new wing-like structure at the outer edge. Comparison with the CCF and CVF waveriders showed that the RF1 and RF2 waveriders had larger volume and higher volumetric efficiency and volume concentration on the symmetric meridian plane. In terms of aerodynamic performance, the values of $C_L$, $C_D$, $L/D$, $|C_M|$, and $X_{cp*}$ of the RF1 and RF2 waveriders were between those of the CCF and CVF waveriders under design and off-design conditions. In short, the RF1 and RF2 waveriders exhibited more-balanced geometric and aerodynamic performances.

The present study also indicates that diverse RFs for the *Ma* distribution can be achieved by adjusting the control parameters so that the waverider would obtain optimized geometric and aerodynamic characteristics.

Capability for WENO Schemes, Journal of Scientific Computing, 2021, 88(3): 75-130.